\documentclass[epj]{svjour}
\usepackage{latexsym}
\usepackage{graphics}
\usepackage{graphicx}  
\usepackage{epsfig} 
\usepackage{subfigure}
\usepackage{lineno}  
\usepackage{amsmath,amssymb}
\usepackage{comment}
\usepackage{multirow}
\usepackage{dcolumn}
\usepackage{bm}
\usepackage{setspace}
\includecomment{printrobustnotes}
\includecomment{printallnotes}
\usepackage{xspace}
\usepackage{color}
\usepackage{seqsplit}
\newcommand{\gevcc}[1]  {\ensuremath{#1~\mathrm{GeV}/c^{2}}}

\def\kforty{$\rm ^{40}K$~}

\def\utwothirtyeight{$\rm ^{238}U$~}
\def\thtwothirtytwo{$\rm ^{232}Th$~}
\def\rntwotwotwo{$\rm ^{222}Rn$~}

\def\pbtwoten{$\rm ^{210}Pb$~}
\def\potwoten{$\rm ^{210}Po$~}
\def\bitwoforteen{$\rm ^{214}Bi$~}

\def\rntwotwenty{$\rm ^{220}Rn$~}

\def\potwoforteen{$\rm ^{214}Po$~}
\def\potwosixteen{$\rm ^{216}Po$~}

\def\naitl{NaI(Tl)~}

\begin{document}
\title{Understanding internal backgrounds of NaI(Tl) crystals toward a 200~kg array for the KIMS-NaI experiment}

\author{
  P.~Adhikari\inst{1} \and
  G.~Adhikari\inst{1} \and
  S.~Choi\inst{2} \and
  C.~Ha\inst{3}\thanks{Corresponding Author : {cha@ibs.re.kr}} \and
  I.S.~Hahn\inst{4} \and
  E.J.~Jeon\inst{3} \and
  H.W.~Joo\inst{2} \and
  W.G.~Kang\inst{3} \and
  H.J.~Kim\inst{5} \and
  H.O.~Kim\inst{3} \and
  K.W.~Kim\inst{2} \and
  N.Y.~Kim\inst{3} \and
  S.K.~Kim\inst{2} \and
  Y.D.~Kim\inst{3,1} \and
  Y.H.~Kim\inst{3,6} \and
  H.S.~Lee\inst{3}\thanks{Corresponding Author : {hyunsulee@ibs.re.kr}} \and
  J.H.~Lee\inst{3} \and
  M.H.~Lee\inst{3} \and
  D.S.~Leonard\inst{3} \and
  J.~Li\inst{3} \and
  S.Y.~Oh\inst{1} \and
  S.L.~Olsen\inst{3} \and
  H.K.~Park\inst{3} \and
  H.S.~Park\inst{6} \and
  K.S.~Park\inst{3} \and
  J.H.~So\inst{3} \and
  Y.S.~Yoon\inst{3}
}                     

\institute{
  Department of Physics, Sejong University, Seoul 05006, Korea \and
  Department of Physics and Astronomy, Seoul National University, Seoul 08826, Korea \and
  Center for Underground Physics, Institute for Basic Science, Daejeon 34047, Korea \and
  Department of Science Education, Ewha Womans University, Seoul 03760, Korea \and 
  Department of Physics, Kyungpook National University, Daegu 41566, Korea \and
  Korea Research Institute of Standards and Science, Daejeon 34113, Korea
}
\date{Received: date / Revised version: date}
%
\abstract{
 The Korea Invisible Mass Search (KIMS) collaboration has developed low-background NaI(Tl) crystals that are suitable for the direct detection of WIMP dark matter. Building on experience accumulated during the KIMS-CsI programs, the KIMS-NaI experiment will consist of a 200~kg NaI(Tl) crystal array surrounded by layers of shielding structures and will be operated at the Yangyang underground laboratory. The goal is to provide an unambiguous test of the DAMA/LIBRA annual modulation signature. Measurements of six prototype crystals show progress in the reduction of internal contamination from radioisotopes. Based on our understanding of these measurements, we expect to achieve a background level in the final detector configuration that is less than 1 count/day/keV/kg for recoil energies around 2~keV. The annual modulation sensitivity for the KIMS-NaI experiment shows that an unambiguous 7$\sigma$ test of the DAMA/LIBRA signature would be possible with a 600~kg$\cdot$year exposure with this system.
} 
\maketitle
\section{Introduction}
The detection of light signals from scintillation crystals is a well established technology for searching for extremely rare events such as weakly interacting massive particles (WIMPs)~\cite{crystalwimp,hslee07,sckim12,bernabei13,cresst2014,znwo4}, a strongly motivated particle dark matter candidate. A WIMP can elastically scatter off a nucleus and a portion of the recoil energy that is deposited in the crystal is transformed into scintillation light. For these searches, it is essential to produce ultra low-background crystals by reducing internal radioisotope contaminations to extremely low levels. Among various crystal scintillators, NaI(Tl) is particularly interesting because of the positive signal reported by the DAMA/LIBRA collaboration for an annual modulation in the rate for low-energy events in an array of NaI(Tl) crystals~\cite{bernabei13,bernabei08,bernabei10}.
This modulation signal  has been the subject of a continuing debate because other experiments observe no modulation~\cite{Aprile:2015ibr,Abe:2015eos} and because the WIMP-nucleon or WIMP-electron cross sections inferred from the DAMA/LIBRA modulation are in conflict with limits from other experiments, such as LUX~\cite{agnese14}, SuperCDMS~\cite{akerib14}, CRESST-II Phase 2~\cite{Angloher:2014myn}, and XENON100~\cite{Aprile:2015ade,aprile12}. However, it is possible to explain all of the direct search experimental results in terms of non-trivial systematic differences in detector responses and possible modifications of the commonly used halo model for the galactic distribution of dark matter~\cite{Arina:2014yna,Bernabei:2014bqa}.

The Korea Invisible Mass Search~(KIMS) experiment searched for WIMP dark matter using ultra low-\seqsplit{back-ground} CsI(Tl) crystal detectors~\cite{kims_pow,kims_pow1,kims_crys,kims_crys1}. Null results from the KIMS-CsI experiment establish stringent constraints on the interpretation of the DAMA/LIBRA annual modulation signal as being due to WIMP-iodine interactions~\cite{hslee07,sckim12}. However, an alternative interpretation of the DAMA/LIBRA result that attributes it to being primarily due to WIMP-sodium nuclei interactions, as would be the case for low-mass WIMPs, cannot be ruled out. Therefore, it remains necessary to try to reproduce the DAMA/LIBRA observations with crystal detectors of the same NaI(Tl) composition.
Efforts by various groups such as DM-Ice~\cite{dmice}, ANAIS~\cite{amare14A}, and SABRE~\cite{Xu:2015wha} to test DAMA/LIBRA signals using the same NaI(Tl) crystals have been initiated.
Recently, the KIMS collaboration engaged in R\&D to produce ultra low-background NaI(Tl) crystals~\cite{kykim15}. The successful growth of crystals with background levels at or below those measured in the DAMA/LIBRA experiment (1~count/day/keV/kg) is required to achieve a stringent and unambiguous test of the DAMA/LIBRA annual modulation observation. 

In this article, we describe internal background measurements of NaI(Tl) crystals at the Yangyang underground laboratory (Y2L). Six R\&D stage crystals were grown by Alpha Spectra Inc.~\footnote{~http://www.alphaspectra.com} and Beijing Hamamatsu~\footnote{~http://www.bhphoton.com} using a variety of NaI powders. Background levels measured in the six crystals permitted us to understand the dominant contributions to the WIMP signal region and provided guidance for the further reduction of these backgrounds. Based on our best understanding of the background, the ultimate sensitivity of a 200~kg array of the KIMS-NaI experiment is determined.

\section{Background measurements of NaI(Tl) crystals}

\subsection{The Experimental Setup}
To evaluate the background contamination levels in the NaI(Tl) crystals, we used the same experimental apparatus that was used for the KIMS-CsI experiment at Y2L~\cite{sckim12,kykim15}. This includes a 12-module array of CsI(Tl) crystals inside a shield that includes 10~cm of copper, 5~cm of polyethylene, 15~cm of lead, and 30~cm of liquid-scintillator-loaded mineral oil. The shielding stops external neutrons and gamma rays, and vetoes cosmic-ray muons. The availability of the already-established setup has been advantageous for a timely evaluation of new crystals.

\begin{figure*}[!htb]
    \begin{tabular}{cc}
      \includegraphics[width=0.9\textwidth]{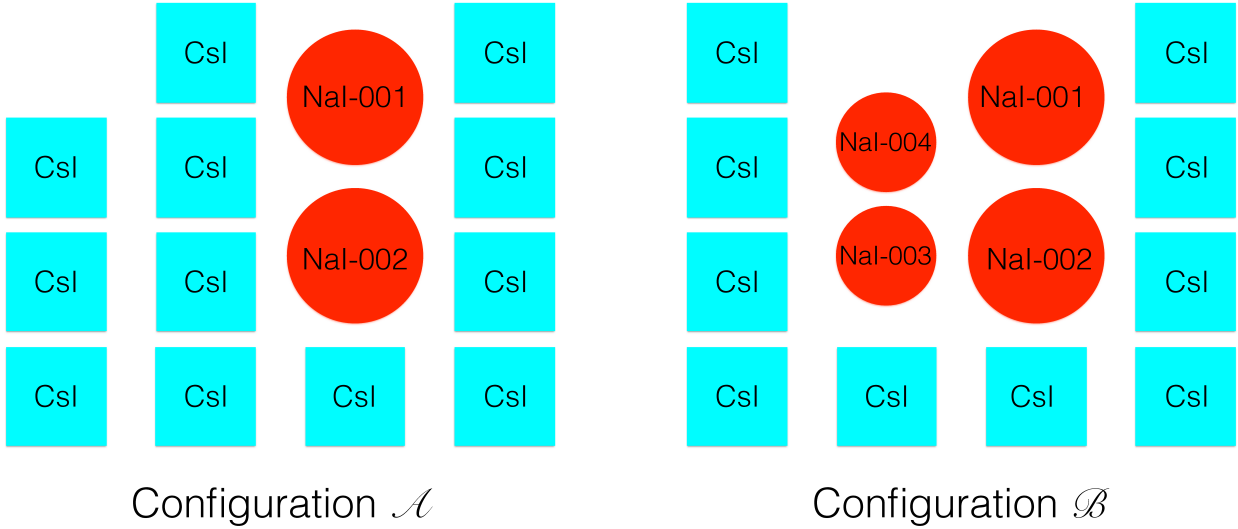}&
    \end{tabular}
    \caption{Schematic drawings of test setups for NaI(Tl) crystals~(circles) with CsI(Tl) crystal array (squares). The first two crystals~(NaI-001 and NaI-002) were first tested with configuration $\mathcal{A}$. In the $\mathcal{B}$ configuration, we tested NaI-003 and NaI-004 crystals together with NaI-001 and NaI-002. Subsequently, NaI-003 and NaI-004 were replaced with NaI-005 and NaI-006.}
  \label{arrayconfig}
\end{figure*}

So far, six R\&D stage NaI(Tl) crystals have been tested inside the CsI(Tl) detector array. From February 2014 to September 2014, two NaI(Tl) crystals (NaI-001 and NaI-002) were tested in the ``$\mathcal{A}$'' configuration (see Fig.~\ref{arrayconfig}). Results from this test were previously reported in Ref.~\cite{kykim15}. In September 2014, two NaI(Tl) crystals~(NaI-003 and NaI-004) were delivered and promptly used to replace two CsI(Tl) crystals as shown in the ``$\mathcal{B}$'' configuration of Fig.~\ref{arrayconfig}. In December 2014 and January 2015, we replaced NaI-003 and NaI-004 with the newly-arrived NaI-005 and NaI-006 crystals. Crystals that arrive hereafter will also be evaluated in this setup.

\subsection{NaI(Tl) crystals}
The low background NaI(Tl) crystals were grown from highly purified powders.
To reach the required background level in the crystal form, it is necessary to study the powder purification procedure and understand the crystal \seqsplit{gr-owth} environment.

Of the six crystals tested, the first five crystals were produced by Alpha-Spectra~(AS). The sixth one was produced by Beijing Hamamatsu (BH). Sigma-Aldrich~(SA) and AS applied several purification procedures to the powders to reduce levels of radionuclides. NaI-001 and NaI-002 were produced by AS from their own powders. The details of the first two crystals are discussed in Ref.~\cite{kykim15}. The NaI-003 crystal, grown from SA Astro-Grade~(AG) powder, is known to contain less contamination from potassium.
NaI-004 and NaI-006 were made of SA Crystal-Growth~(CG) grade powder but were grown separately by two different crystal growers. For the NaI-005 production, AS developed additional background reduction methods, especially aimed at reducing \pbtwoten contamination in the powder and crystal growing process.

The crystals are cylindrical in shape and were cut from large ingots. After the crystals were polished in a pure nitrogen environment by a proprietary process they were wrapped with a Teflon reflector and inserted into an oxygen-free copper case. The circular plane of the crystal was light-coupled with a quartz window via a gel-type optical pad. Each end of the encapsulated cylinder was attached to a 3-inch diameter, glass window R12669 photomultiplier tube~(PMT) manufactured by Hamamatsu Photonics~\footnote{~http://www.hamamatsu.com}.
The specifications of the crystals are summarized in Table~\ref{crystalspec}.
\begin{table*}
\begin{center}
\caption{Specification of the NaI(Tl) crystals. The second column specifies the cylinder diameters and length in inches. The ``Powder'' acronyms are AS-B (AS-C): Alpha Spectra purified powder, SA-AG: Sigma-Aldrich Astro-Grade powder (reduced K content), SA-CG: Sigma-Aldrich Crystal-Growth powder, and AS-WSII: Alpha Spectra WIMPScint-II grade powder. The names of the crystal growers are AS: Alpha-Spectra and BH: Beijing-Hamamatsu. The last two columns are the year and month the crystals were grown and transported to Y2L.}
\label{crystalspec}
\begin{tabular}{lcccccc}
  Crystal   & Size (D$\times$L)        & Mass(kg)  & Powder &    Crystal Grower  & T (Growth) & T (Y2L) \\ \hline
  NaI-001  &   5$''$ $\times$ 7$''$    & 8.3  &  AS-B    & AS &    2011.9  & 2013.9 \\
  NaI-002  & 4.2$''$ $\times$ 11$''$   & 9.2  &  AS-C    & AS &    2013.4  & 2014.1 \\
  NaI-003  & 4.5$''$ $\times$ 3.5$''$  & 3.3  &  SA-AG & AS & 2014.4     & 2014.9\\
  NaI-004  & 4.5$''$ $\times$ 3.5$''$  & 3.3  &  SA-CG  & AS & 2014.3    & 2014.9\\
  NaI-005  & 4.2$''$ $\times$ 11$''$   & 9.2  &  AS-WSII & AS & 2014.7   & 2014.12\\
  NaI-006  & 4.8$''$ $\times$ 8.8$''$  & 11.4 &  SA-CG & BH& 2014.10     & 2015.1\\
\end{tabular}
\end{center}
\end{table*}

\subsection{Internal Natural Backgrounds}
To produce ultra low-background NaI(Tl) crystals, one should understand internal contamination of natural radioisotopes. Table~\ref{internalbackgrounds} lists the contamination levels for the six crystals as measured with the test configurations described above.
\begin{table*}[ht]
\begin{center}
\caption{Background rates from internal radioactive contaminants in the NaI(Tl) crystals. For the ``$\alpha$ Rate'', each alpha particle is counted as one decay. The light yields are measured using 59.6~keV gamma events from $^{241}$Am calibration source. A chain equilibrium is assumed for the interpretation of radioactivity measurements related to $^{238}$U and $^{232}$Th, with the exception of $^{210}$Pb.}
\label{internalbackgrounds}
\begin{tabular}{lccccccc}
  Crystal  & $^{\rm nat}$K (\kforty)   & $^{\rm nat}$K (ICP-MS)  & \utwothirtyeight & \thtwothirtytwo & $\alpha$ Rate& Light Yield        \\
  (unit)   & (ppb)        & powder (ppb)    & (ppt)            & (ppt)           &  (mBq/kg)                      & (Photoelectrons/keV)           \\
  \hline                                                                                           
  NaI-001  & $40.4\pm2.9$ &  $-$ &  $<0.02$         & $<3.2$         & $3.29\pm0.01$   &  $15.6\pm1.4$    \\
  NaI-002  & $48.1\pm2.3$ &  $-$ &  $<0.12$         & $0.5\pm0.3$     & $1.77\pm0.01$   &  $15.5\pm1.4$   \\
  NaI-003  & $25.3\pm3.6$ &  $25$ &  $<0.14$         & $0.5\pm0.1$   & $2.43\pm0.01$   &  $13.3\pm1.3$     \\
  NaI-004  & $>116.7$     &  $100-400$     &  $-$             & $-$             & $-$             &  $3.9\pm0.4$     \\
  NaI-005  & $40.1\pm4.2$ &  $50$ &  $<0.04$         & $0.2\pm0.1$  & $0.48\pm0.01$    & $12.1\pm1.1$     \\
  NaI-006  & $>127.1$     &  $100-400$ &  $<0.05$         & $8.9\pm0.1$    & $1.53\pm0.01$   &  $4.4\pm0.4$     \\
\end{tabular}
\end{center}
\end{table*}

\subsubsection{\kforty\  background}
One of the most dangerous sources of background for WIMP searches with \naitl crystals is contamination from $^{40}$K.
Its natural abundance is roughly 0.012\% of the total \seqsplit{a-mount} of potassium ($^{\rm nat}$K).
About 10\% of \kforty\ decays produce a 1460~keV $\gamma$-ray in coincidence with a 3~keV $X$-ray. If the 1460~keV $\gamma$-ray escapes the crystal and only the 3~keV $X$-ray is detected, an event is produced that is similar in energy to that expected for a WIMP-nuclei interaction~\cite{Lewin:1995rx}. The $^{\rm nat}$K contents in the DAMA crystals
are in the 10$-$20~ppb range~\cite{bernabei08had}. 

In our apparatus, \kforty decays can be identified by coincidence signals between 1460~keV $\gamma$-rays in the CsI(Tl) detectors and 3~keV $X$-rays in the NaI(Tl). The \kforty\ background level in each crystal is determined by comparing the measured coincidence rate with a Geant4-simulated rate using the method described in Ref.~\cite{kykim15}. To understand the origin of the \kforty background, $^{\rm nat}$K contents in powders used in the crystal production are measured by inductively coupled plasma mass spectrometry (ICP-MS) and the results are compared to the crystal measurements.

In three of the crystals~(NaI-001, NaI-002, and NaI-005) produced by the AS company, the coincidence analysis results
show $^{\rm nat}$K levels in the range of 40$-$50~ppb.
For the AS WIMPScint-II grade (AS-WSII) powder, which was used to grow the NaI-005 crystal, ICP-MS measurements similarly
show $^{\rm nat}$K contamination of about 50~ppb.

The SA powder comes in two varieties, each with different levels of $^{\rm nat}$K contamination. The SA-CG powder is commonly used for commercial NaI(Tl) crystals while the SA-AG powder developed for rare event searches shows reduced contamination of radioisotopes. Using its ICP-MS measurements, the SA company reported that the SA-CG powders contain $^{\rm nat}$K levels of approximately 100$-$400~ppb depending on the batch, while the SA-AG powders have batch-to-batch variation $^{\rm nat}$K levels, all in the range of 3$-$30~ppb.
The NaI-003 crystal that was grown from the SA-AG powder has a measured $^{\rm nat}$K level of $25.3\pm3.6$~ppb. This is the lowest $^{\rm nat}$K contamination of the six tested crystals, and is in good agreement with the vendor's ICP-MS measurement of 25~ppb for this batch.
Due to lower light yield from NaI-004 and NaI-006, which were made with SA-CG powder, our measurements could only establish a limit on the $^{\rm nat}$K contamination level. The $^{\rm nat}$K level in these crystals was measured to be more than 100~ppb, consistent with contamination levels in the SA-CG powder.

Based on these measurements, we conclude that the source of \kforty contamination in the NaI(Tl) crystals originates mostly from contamination in the NaI powder, and that no significant additional contamination is introduced during the crystal growing procedure.
The $^{\rm nat}$K measurements for the six crystals are listed in Table~\ref{internalbackgrounds}.

For further reduction of the \kforty\ contamination, several approaches are currently under development. First, WIMPScint-III grade powder is produced by the AS company with lower $^{\rm nat}$K contamination than that of AS-WSII. Several ICP-MS measurements on this powder show on average of 16~ppb $^{\rm nat}$K contamination. AS is currently growing a crystal with this powder.
Separately, we have acquired 10~ppb SA-AG powder which is also in the crystal growing process.
Second, the simultaneous production of a 1460~keV $\gamma$-ray with a 3~keV $X$-ray makes it possible to tag these events with surrounding detection devices. In the current setup, a Geant4-based simulation compared with data shows that the surrounding CsI(Tl) crystals tag 40\% of $^{40}$K-induced events. Studies to increase the veto-tagging efficiency by incorporating the crystals inside a liquid scintillator veto~(LSV) system are being performed. A separate Geant4 simulation estimates that the tagging efficiency with a 40~cm thick LSV system can be better than 80\%. Therefore, the implementation of such a system for the KIMS-NaI experiment is expected to reduce the $^{40}$K-induced background by more than a factor of two.
Finally, R\&D on procedures for additional potassium reduction in the NaI powder are in progress. Column chromatography with certain resin compounds is known to remove contaminants in solutions via an ion-exchange mechanism. In the near future, we expect that some combination of all of the above will reduce the potassium background to a level lower than the level that has been achieved by DAMA/LIBRA.

\subsubsection{\utwothirtyeight\ background}
\utwothirtyeight is one of the most common radioisotopes in nature primarily because of its long decay time. Assuming an equilibrium in the \utwothirtyeight decay chain, the contamination level in NaI(Tl) crystals can be evaluated by exploiting the 237~$\mu$s mean lifetime of a \potwoforteen\ $\alpha$-decay (E$_{\alpha}$=7.7~MeV), which immediately follows its production via $\beta$-decay of \bitwoforteen as described in Refs.~\cite{kims_crys,kykim15}. Taking advantage of pulse shape differences between $\alpha$-induced and $\beta$-induced events, we used this technique to determine \utwothirtyeight contamination levels in the six crystals.
We found no significant time-\seqsplit{dep-endent} exponential components that are a characteristic of $^{214}$Po.
Upper limits on the activities were determined; no limit could be set for NaI-004 because of its low light yield. The $^{238}$U background levels in the crystals are all below the ppt level (see Table~\ref{internalbackgrounds}).

\subsubsection{\thtwothirtytwo\ background}
Contamination from the \thtwothirtytwo decay chain was studied by using $\alpha-\alpha$ time interval measurements in the crystals. In this case, we looked for a \potwosixteen\ $\alpha$-decay (E$_{\alpha}$=6.8~MeV) component with a mean decay time of 209~ms that follows its production via $\alpha$-decay of \rntwotwenty(E$_{\alpha}$=6.3~MeV). Figure~2 shows the distributions of the time differences with exponential fits between two adjacent $\alpha$-induced events for NaI-005 and NaI-006. Other than NaI-006, all crystals show a small exponential component below 1~s for the characteristic \potwosixteen\ decay time that indicates a small amount of \thtwothirtytwo contamination, assuming a chain equilibrium. NaI-006 contains a \potwosixteen\ decay component, corresponding to a $8.9\pm0.04$~ppt of the $^{232}$Th contamination. 
The $^{232}$Th contents in all of the crystals, inferred assuming throughout the $^{232}$Th decay chain, are listed in Table~\ref{internalbackgrounds}. 
\begin{figure*}[!htb]
  \begin{center}
    \begin{tabular}{cc}
      \includegraphics[width=0.45\textwidth]{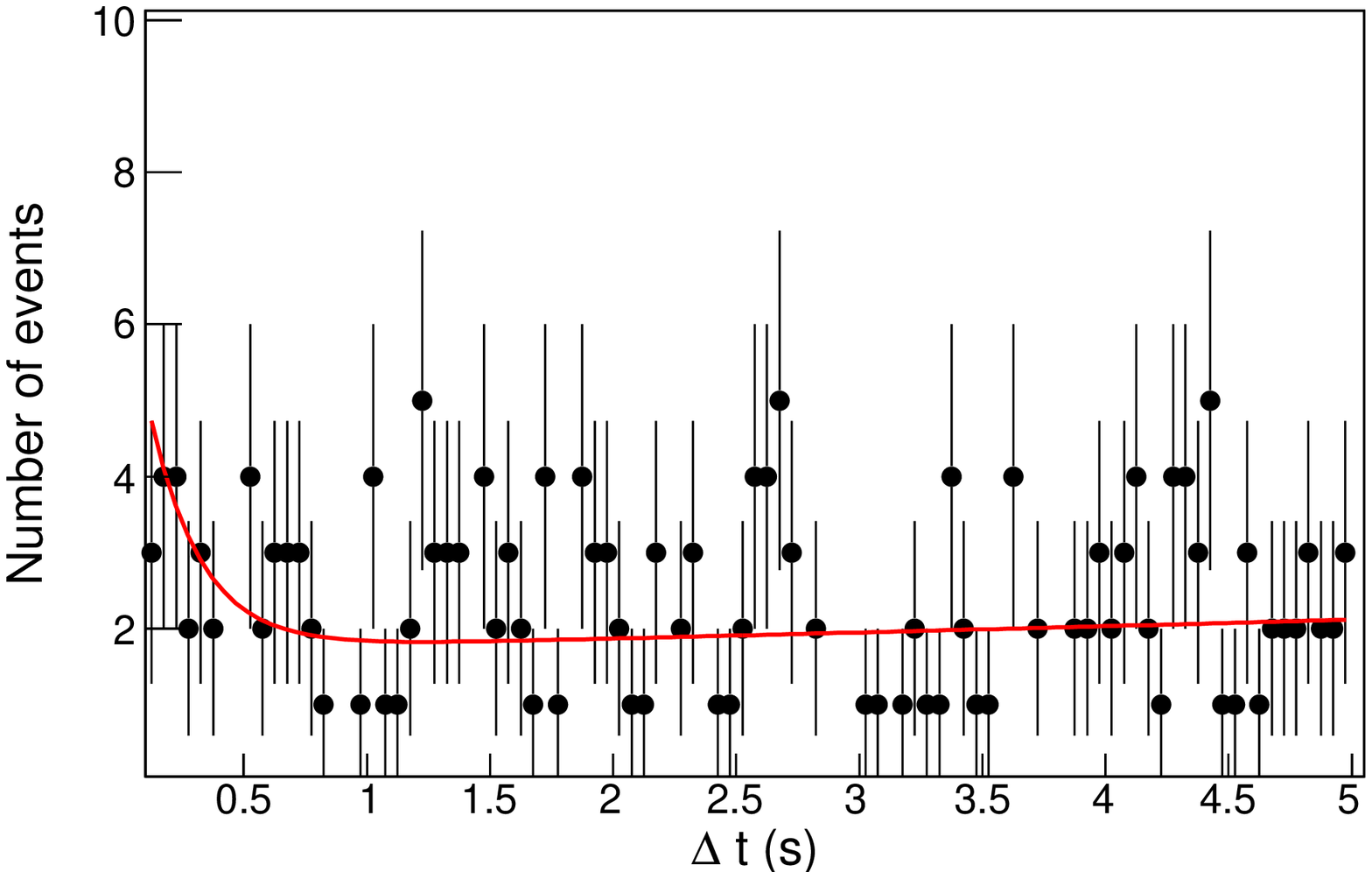} &
      \includegraphics[width=0.45\textwidth]{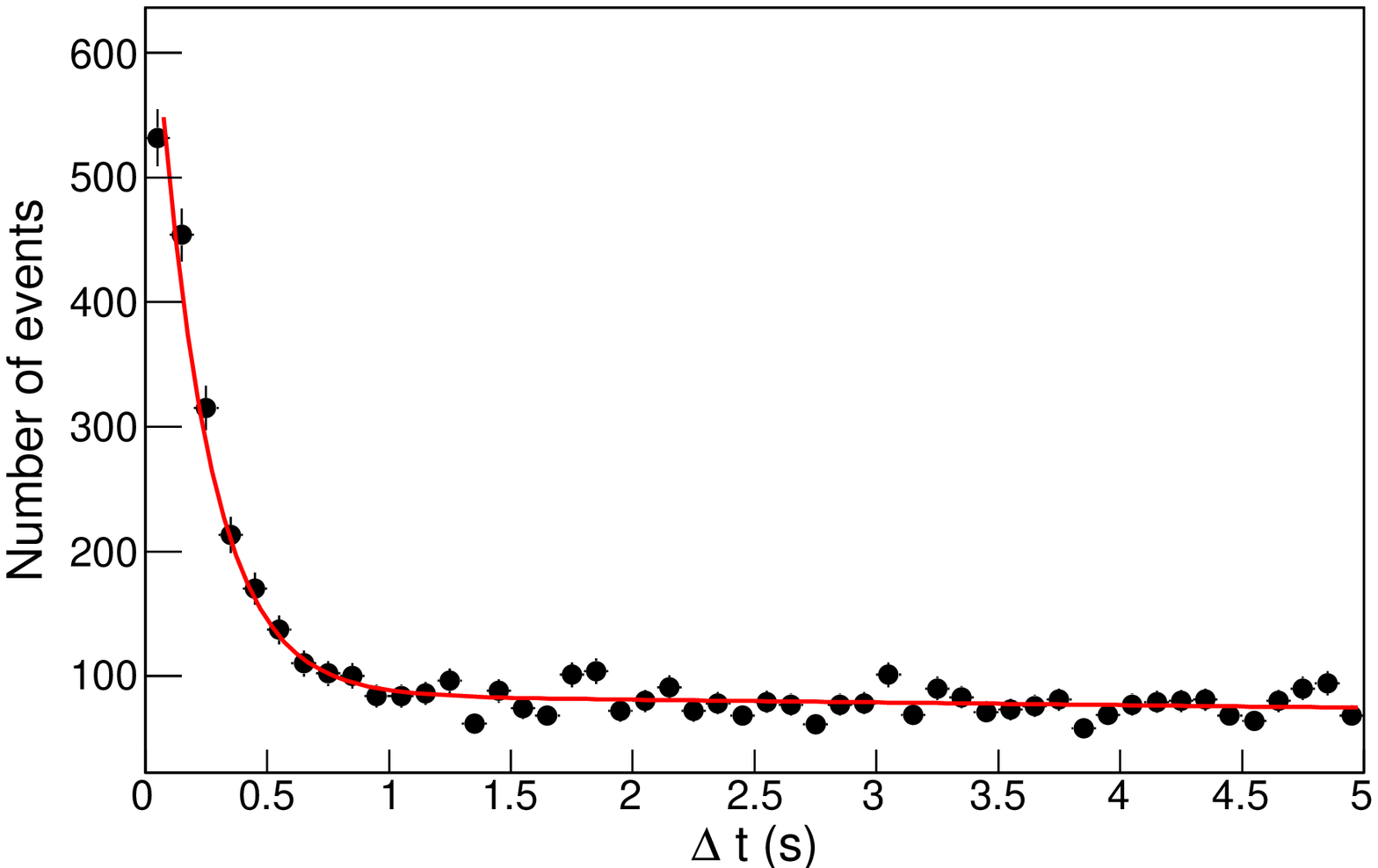}\\
      (a) NaI-005 & (b) NaI-006\\
    \end{tabular}
    \caption{The time difference distribution and exponential fits between two successive $\alpha$-induced events in the (a)~NaI-005 and (b)~NaI-006 crystals. The exponential components below 1~s are due to \rntwotwenty and \potwosixteen sequential decays. While the fit for NaI-005 is consistent with a flat line, the fit for NaI-006 shows these decay components indicating $^{232}$Th contamination.}
  \end{center}
  \label{coincidence}
\end{figure*}

\subsubsection{\pbtwoten\ background}
The levels of \utwothirtyeight\ and \thtwothirtytwo\ contamination measured in the six crystals are too low to account for the total observed $\alpha$ rates.  This suggests that the bulk of the $\alpha$ rate is due to decays of \potwoten(E$_{\alpha}$=5.4~MeV) nuclei originating from \rntwotwotwo\ exposure that occurred sometime during the powder and/or crystal processing stages. This was confirmed by the observation of a 46.5~keV $\gamma$ peak, which is characteristic of $^{210}$Pb. The \potwoten level in NaI-005 was estimated from the total alpha yield to be about 1/3 of that in NaI-002 (see Table~\ref{internalbackgrounds}), consistent with the 46.5~keV $\gamma$-ray peak measurements.

\begin{figure*}[!htb]
\begin{tabular}{cc}
\includegraphics[width=0.45\textwidth]{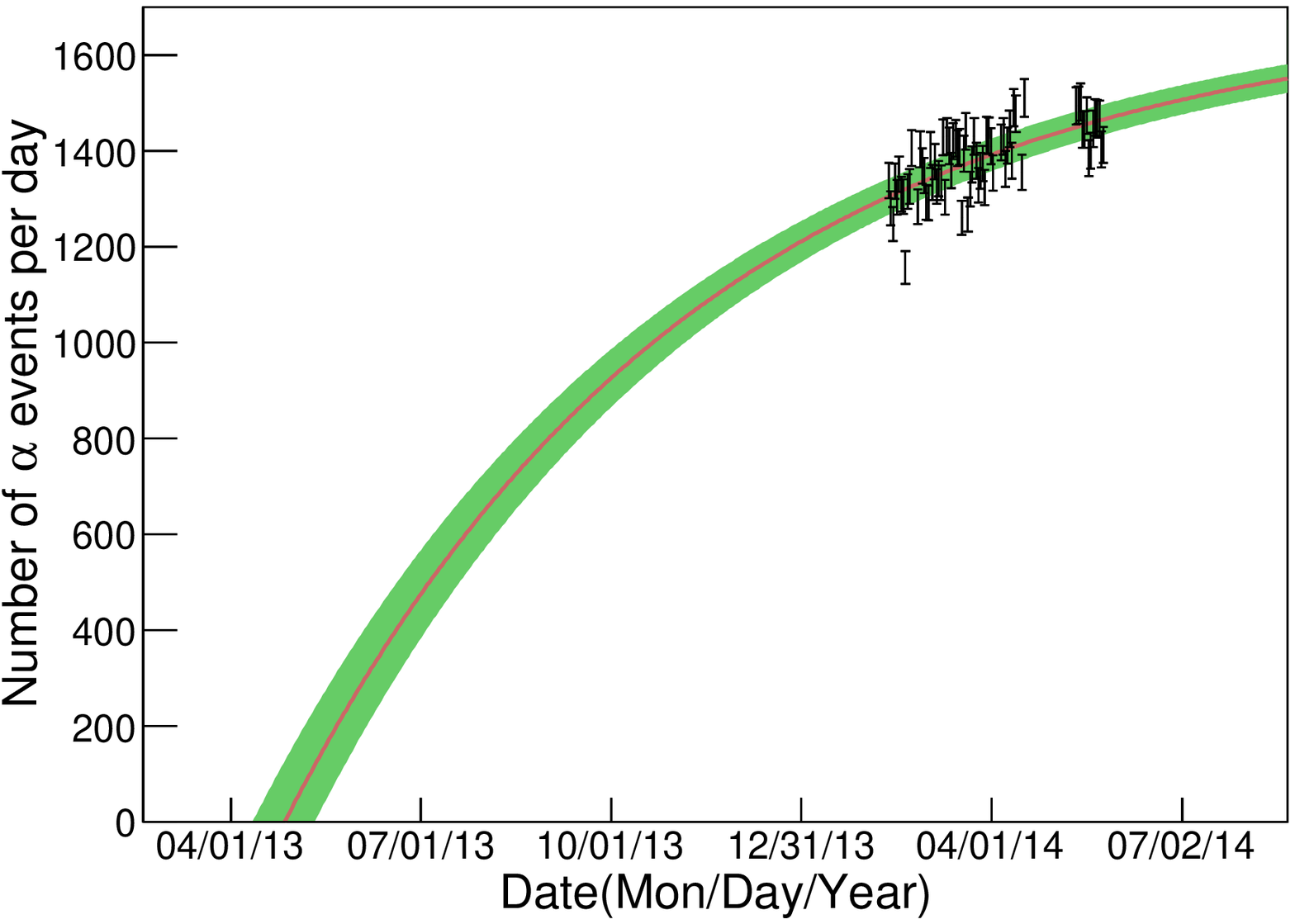}&
\includegraphics[width=0.45\textwidth]{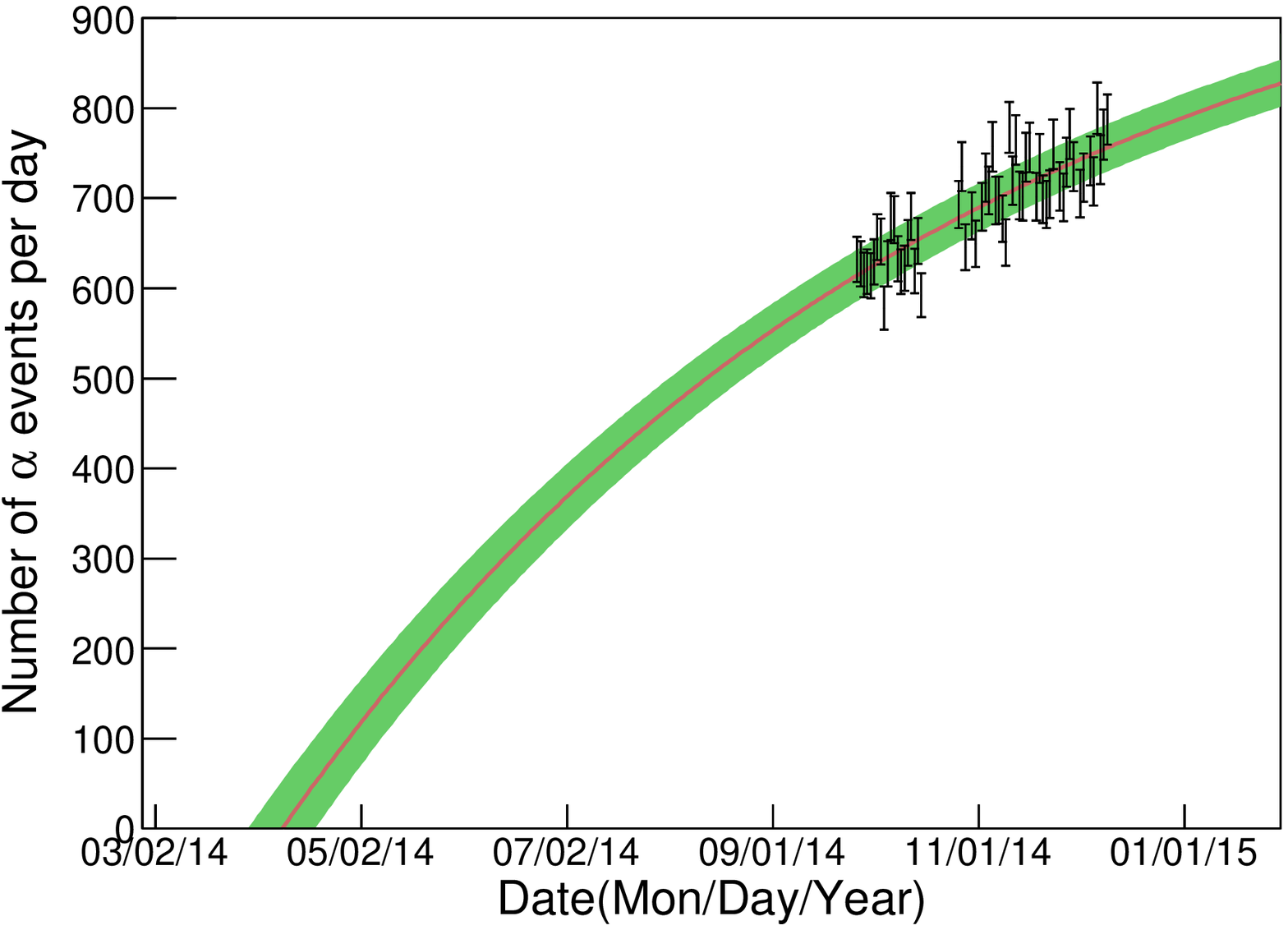}\\
(a) NaI-002  & (b) NaI-003  \\
\includegraphics[width=0.45\textwidth]{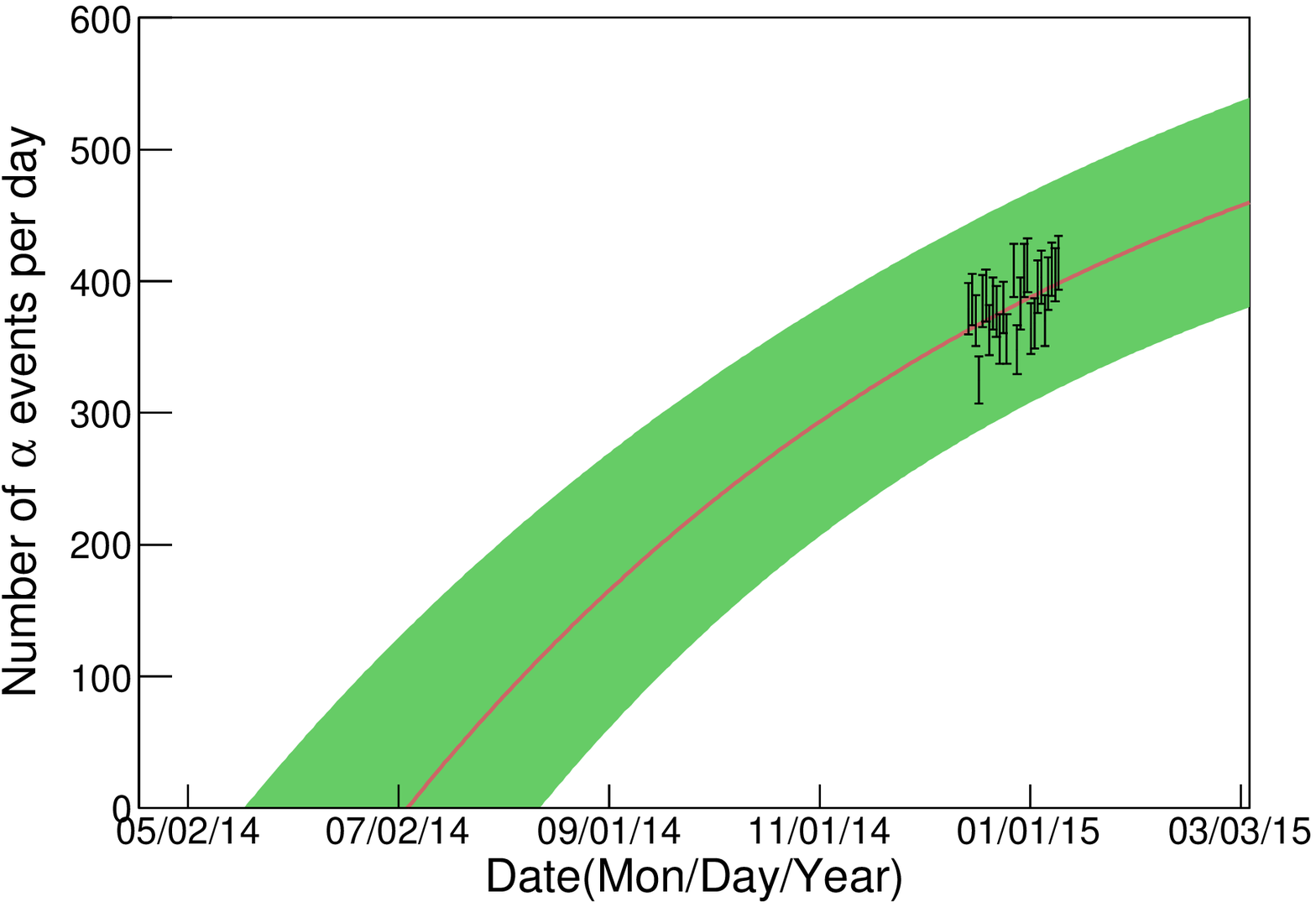}&
\includegraphics[width=0.45\textwidth]{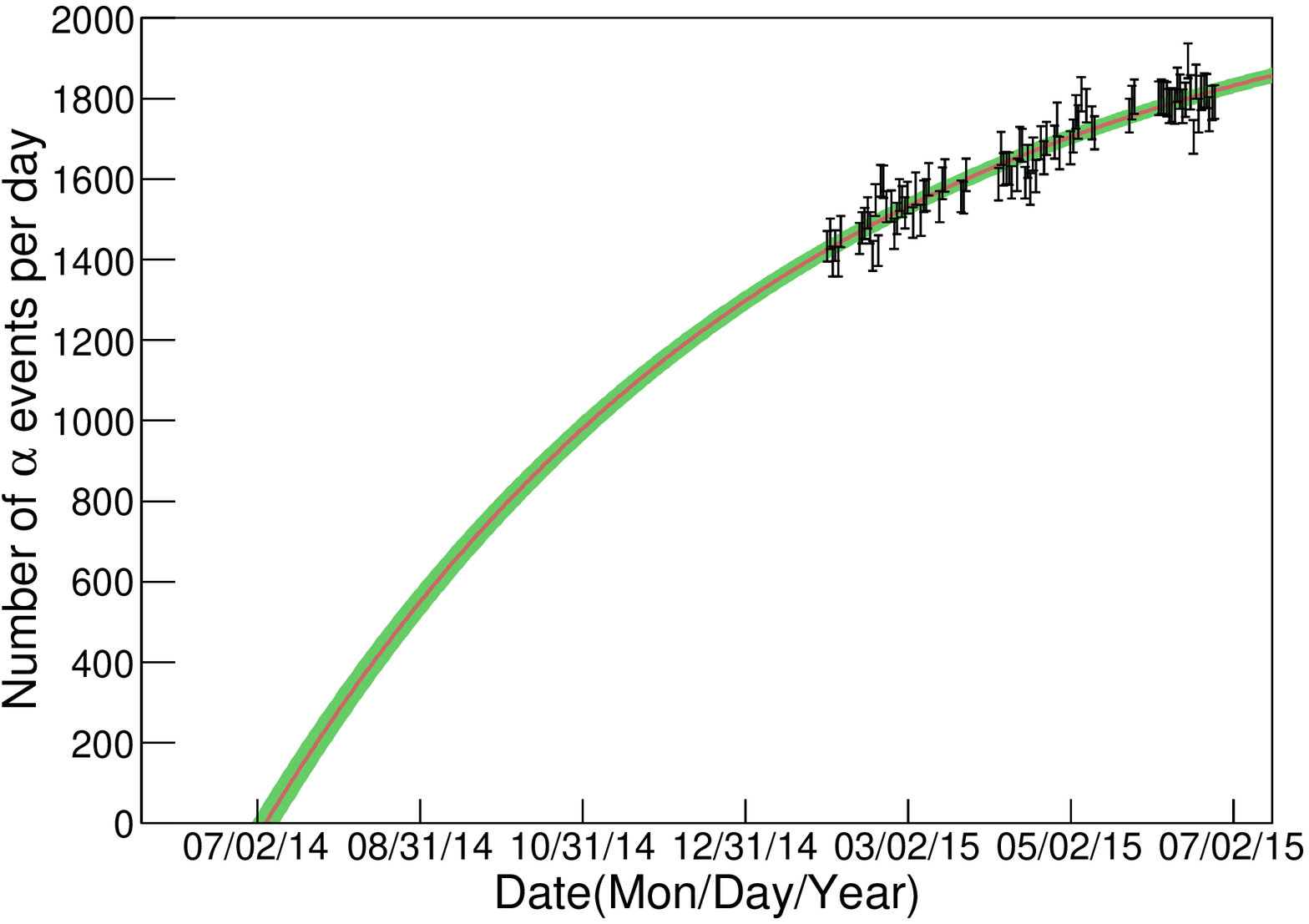}\\
(c) NaI-005 & (d) NaI-006 \\
\end{tabular}
\caption{The increase in alpha activities is fitted to a model that assumes a single, instantaneous \potwoten contamination event occurred for each crystal: (a)~NaI-002, (b)~NaI-003, (c)~NaI-005, and (d)~NaI-006. The best fit is indicated as the red line with its 68\% uncertainty as a green band. 
  Since the event sample in the NaI-006 crystal contains additional $\alpha$ events from the substantial \thtwothirtytwo contamination, the best fit date precedes the crystal growth date by roughly three months.
}
\label{alpharatesdaqtime2}
\end{figure*}

In the $^{222}$Rn decay chain, $^{210}$Po ( half life = 138 days) is produced by the beta decay of $^{210}$Pb (half life = 22 years) and $^{210}$Bi (half life = 5 days), and subsequently decays via alpha emission to $^{206}$Pb. Thus, if the $^{210}$Po contamination is zero at the time when $^{222}$Rn exposure occurs, it will grow with a characteristic time of $\tau_{\rm Po}=200$ days until an equilibrium is reached. This change in the total alpha rate over time can provide information about the date of the \rntwotwotwo\ exposure. Figure~\ref{alpharatesdaqtime2} shows the total alpha rates in the four crystals as a function of data-taking time (days).  
The \potwoten alpha activity grows as 
\begin{equation}  
R_{\alpha}(t)\approx A(1-e^{-(t-t_{0})/ \tau_{\rm{Po}}}),
\end{equation}  
where $t_{0}$ is the
time when the initial \pbtwoten\ contamination occurred, assuming that the contamination 
occurred over a single short time period.
The fits to the time dependence of the $^{210}$Po alpha signals shown in Fig.~\ref{alpharatesdaqtime2} indicate that $t_0$ coincides with the time that the crystal was grown. However, since the preparation of the NaI powder happens about two months prior to the crystal growing, it is still possible that some contamination occurred during powder production. 

We asked the crystal growers to apply more systematic controls on contamination during the crystal growth. NaI-005 was a result of reduced radon contact and additional  purification of the NaI powder by AS.
As listed in Table~\ref{internalbackgrounds}, the total $\alpha$ rate of NaI-005 was reduced by more than a factor of three compared to NaI-002. 
For further reduction of the $^{210}$Pb contamination, we are investigating purification methods for the NaI powder that utilize ion-exchange resins that were previously applied in the Kamland-PICO experiment~\cite{kamlandpico}.

\subsubsection{Background summary and its possible reduction}
\begin{figure}[!htb]
\begin{center}
\includegraphics[width=1.0\columnwidth]{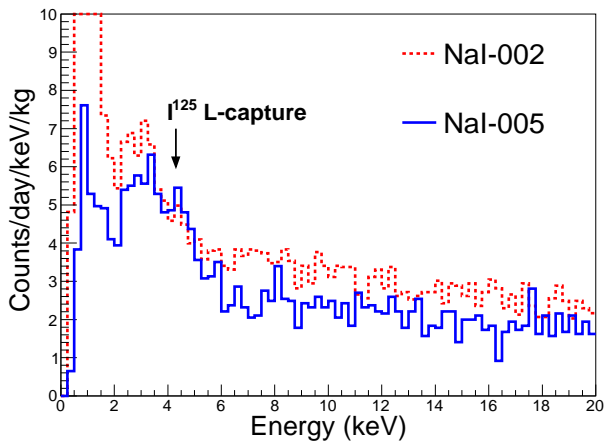}
\end{center}
\caption{Background event spectra at the energies below 20~keV in the NaI-002~(red line) and NaI-005~(blue line) crystals. Note that cosmogenically activated I$^{125}$ peak at around 4-keV is indicated with an arrow for NaI-005 (see text).}
\label{backgroundlevel2}
\end{figure}
Figure~\ref{backgroundlevel2} shows a comparison of the background levels in the NaI-002 and NaI-005 crystals. The factor of three reduction of \pbtwoten in NaI-005 compared to that for NaI-002 results in a 2~dru (differential rate unit = counts/day/keV/kg) background level at 6~keV. Below 6~keV, the main contributors to the remaining background are three types. The $X$-ray of the $^{40}$K decay populates at around 3~keV region. Cosmogenically activated $^{125}$I (T$_{1/2}$=59.4~days) converts via the excited state of $^{125}$Te to the ground state of $^{125}$Te by L-capture, producing $X$-ray/Auger electrons with energies at 4~keV. Similarly, $^{22}$Na (T$_{1/2}$=2.6~years) conversion to $^{22}$Ne produces a 0.85 keV $X$-ray by K-capture. The cosmogenically activated background is expected to disappear after a year underground~\cite{Amare:2014bea}. Because our required background level is less than 1~dru over the 2$-$10~keV energy range, additional reduction of these backgrounds is still needed.

To quantify the background contributions from internal radioisotopes, we used Geant4-based Monte-Carlo simulations. The background contribution from a 40~ppb contamination of $^{\rm nat}$K in NaI-005 produces a background count rate of approximately 0.7~dru in the 2$-$4~keV energy range, of which the possibility for a reduction of an additional factor of nearly two has been demonstrated with the SA-AG powder in NaI-003~(25~ppb). Another factor of two reduction using a different batch of an SA-AG powder (10~ppb) or using an AS-WSIII powder (16~ppb) is foreseen for later crystals. The implementation of the LSV can further reduce this background contribution. 

As for the \pbtwoten contamination in the crystals, the contribution to the total background in NaI-005 is about 0.5~dru for energies below 10~keV. The development of techniques using ion-exchange resins to remove \pbtwoten from NaI powder could lead to an additional reduction of the \pbtwoten background by a factor of two~(to less than 0.3~dru).

The remaining background comes from external sources such as long-lived radioisotopes in the PMTs, lead shields and surrounding rocks, as well as from cosmogenic nuclei, including those from cosmic-ray muon-induced events. The planned LSV system and other veto techniques can tag and veto more than 80\% of the background contributions from these external sources. We expect to achieve 0.5~dru for the external background below 10~keV. Crystal growing with the incorporation of all recent developments is proceeding.
With a subset of the improvements near completion, we expect to reach a background level of less than 1~dru
within a short time scale.

\section{KIMS-NaI experiment with a 200~kg NaI(Tl) array}
Following up on the success of the background reduction for the NaI(Tl) crystals, we plan to grow 200~kg crystals for the KIMS-NaI experiment, which would be composed of a 4$\times$4 array of 12.5~kg NaI(Tl) modules, and could be completed by mid 2016.
We set background goals in the NaI(Tl) crystals of approximately 1~dru in the 2$-$10~keV energy range.
The underground area at Y2L has been expanded, and space has been prepared for this experiment. We are constructing a passive shield consisting of copper, lead, and polyethylene, as well as a plastic-scintillator muon veto and an LSV system. 
At least three years of a stable operation will collect enough data to study the annual modulation and test the DAMA/LIBRA observation.
Here we report sensitivity estimates for the KIMS-NaI experiment and compare them with the DAMA/LIBRA observed signals.

\begin{figure*}[!htb]
    \begin{tabular}{cc}
      \includegraphics[width=0.45\textwidth]{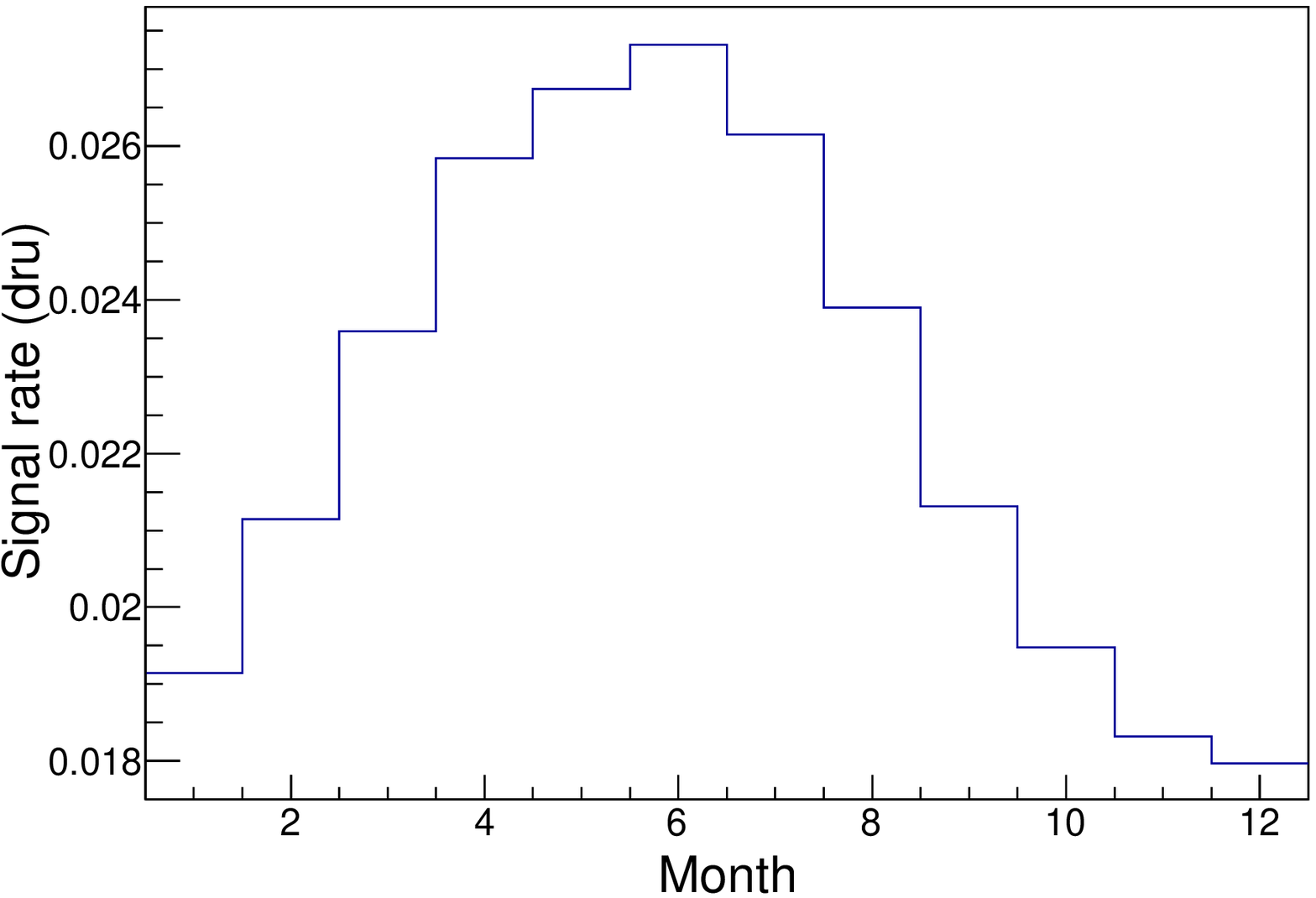}&
      \includegraphics[width=0.45\textwidth]{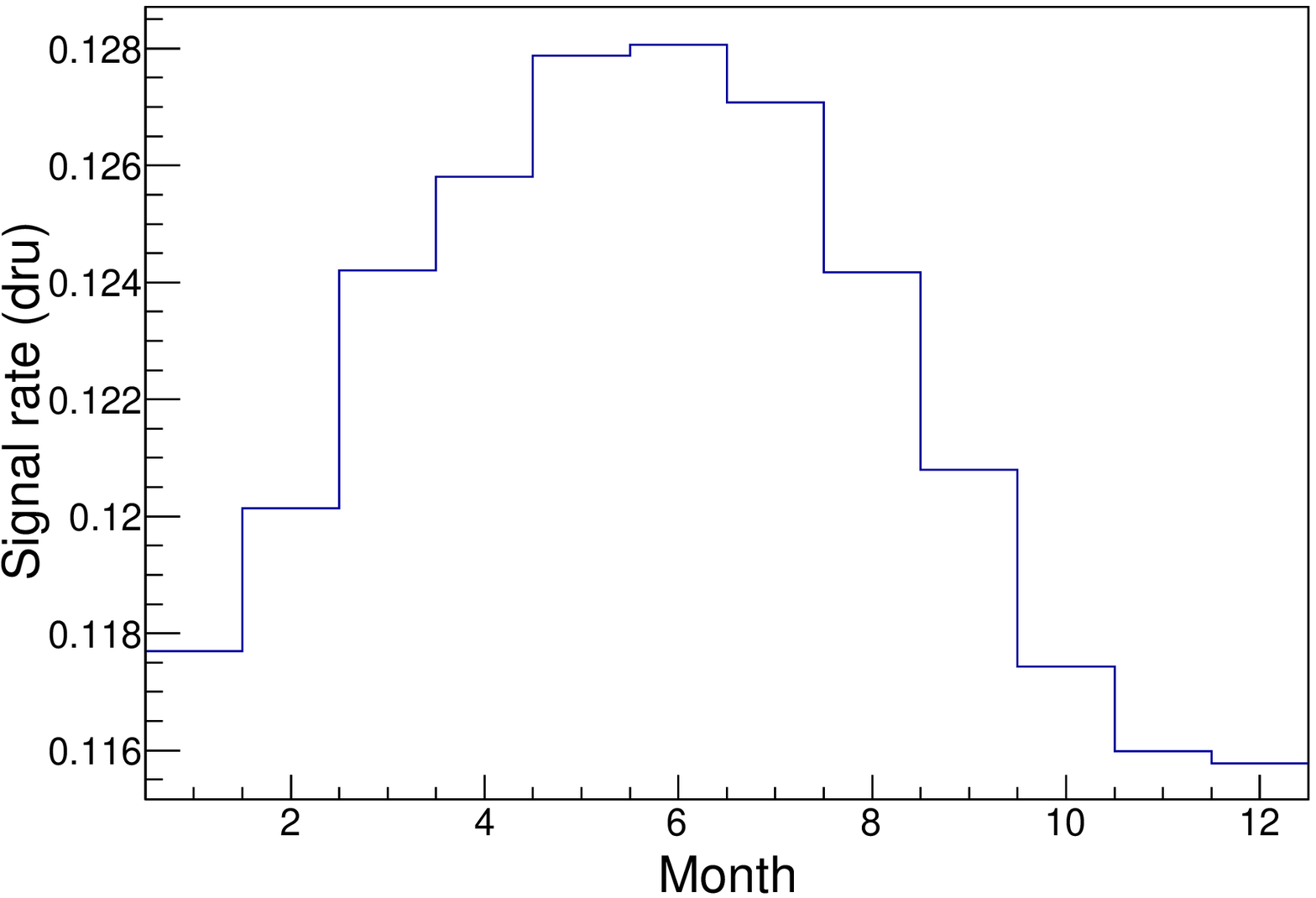}\\
      (a)   &   (b)  \\
    \end{tabular}
    \caption{Annual modulation signals from WIMP-NaI interactions assuming a WIMP mass of \gevcc{10}, a WIMP-nucleon spin-independent cross-section of 3$\times$10$^{-4}$pb (a), a WIMP mass of \gevcc{70}, and a cross-section of 6$\times$10$^{-6}$pb (b) in the energy range between 2 and 6~keV. Scenarios (a) and (b) correspond respectively to WIMP-Na and WIMP-I interpretations of the DAMA/LIBRA annual modulation signals.  
    }
  \label{signal_dama}
\end{figure*}

We generate monthly-binned histograms for WIMP-nuclei scattering signals assuming the standard halo model with the consideration for the Earth's orbital motion around the Sun as discussed in Ref.~\cite{Lewin:1995rx}. In Ref.~\cite{savage09}, the \seqsplit{DAMA/LIBRA} modulation signal is interpreted as a WIMP-nucleon cross section. To make a direct comparison with this interpretation we used the same quenching factors of 0.3 and 0.09 for Na and I, respectively, where the quenching factor is the ratio of the recoil energy deposited by a WIMP to the energy deposited by an electron (or a $\gamma$-ray). Figure~\ref{signal_dama} shows the expected annual modulation signal in the 2$-$6~keV energy range for a \gevcc{10} WIMP and a \gevcc{70} WIMP for WIMP-Na and WIMP-I interactions, respectively. Here, we choose the WIMP-nucleon interaction cross sections that match the modulation amplitude observed by the DAMA/LIBRA experiment. 
For background events, we assume a flat background with no modulation or decay components. We consider a three-year period of uninterrupted data taking with a 200~kg NaI(Tl) crystal array with a 1~dru background rate.

We use an ensemble of Monte Carlo experiments to estimate the sensitivity of the KIMS-NaI experiment, expressed as the expected cross section limits for the WIMP-nucleon spin-independent interactions, in case of no signal. For each experiment, we determine a simulated modulation distribution for a background-only hypothesis. The background rate includes Poisson fluctuations.
We then fit the modulation distribution with a signal-plus-background hypothesis with flat priors for the signal. Additionally, we include a 10\% Gaussian prior for the background. This represents any constraints we may have on the background rate at the time of fitting. This additional constraint has no significant impact on the modulation sensitivity. It is, however, used as a conservative assumption to derive potential sensitivity for WIMP masses outside of the DAMA signal region, where total rate constraints will be stronger than modulation constraints.

\begin{figure}[!htb]
      \includegraphics[width=1.0\columnwidth]{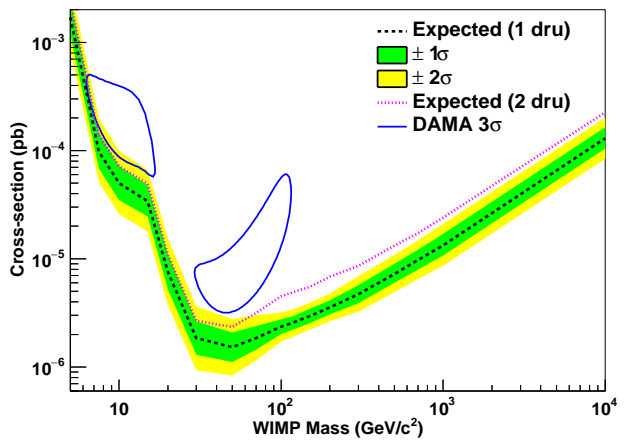}
    \caption{ The median expected 90\% CL upper limit on the WIMP-nucleon spin-independent cross-section assuming the background-only hypothesis (black dotted line), shown together with WIMP-induced DAMA/LIBRA 3$\sigma$-allowed region (solid contour). The dark (green) and light (yellow) shaded bands indicate the 1$\sigma$ and 2$\sigma$ standard deviation probability regions over which the limits have fluctuated. The red dotted line indicates the median expected limit assuming a 2~dru background level. 
    }
  \label{limit}
\end{figure}

The global fit is performed with a Bayesian approach using the Bayesian Analysis Toolkit~\cite{Caldwell:2008fw}. We construct a Bayesian likelihood for each bin (1~keV). The overall likelihood is obtained by multiplying the likelihood for each bin between 2 and 10~keV, considering the energy-dependent rate of WIMP-NaI interactions for WIMP masses between \gevcc{5} and 10~TeV/$c^2$. The 90\% confidence limit~(CL) for each simulated experiment is determined such that 90\% of the posterior density of the WIMP-nucleon cross-section falls below the limit. The median expected 90\% CL limits and their 1$\sigma$ and 2$\sigma$ standard-deviation probability regions are calculated from 2,000 simulated experiments. Figure~\ref{limit} shows the expected median limit, and 1$\sigma$ and 2$\sigma$ standard deviation probability regions, for the KIMS-NaI experiment using the annual-modulation analysis. We also include a more conservative limit that assumes a 2~dru background level, which we already achieved in the R\&D stage crystal. These limits are compared with the DAMA/LIBRA 3$\sigma$ allowed signal region as interpreted in Ref.~\cite{savage09}. Since DAMA/LIBRA reports high significances of the annual modulation signal and its allowed region for the interpreted cross-sections are tightly bounded, KIMS-NaI can exclude or confirm the parameter space with the baseline exposure. As one can see in the figure, three years of KIMS-NaI experimental data can fully cover the DAMA/LIBRA 3$\sigma$ region even for the conservative background assumption of 2~dru.

\begin{figure}[!htb]
    \includegraphics[width=1.0\columnwidth]{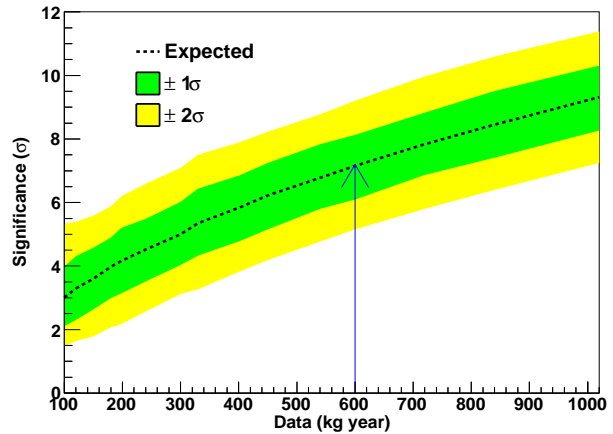}
    \caption{The median expected (black dotted line) significance to observe the same annual modulation signal of DAMA/LIBRA as a function of data size. The dark (green) and light (yellow) shaded bands indicate the 1$\sigma$ and 2$\sigma$ standard deviation probability regions in which the significances are fluctuated. The arrow indicates a three-year data exposure of the KIMS-NaI experiment. 
    }
  \label{significance}
\end{figure}

We also estimate the precision with which we can measure the annual modulation signal if the KIMS-NaI experiment detects a modulation amplitude that is the same as DAMA/LIBRA's. The annual modulation amplitude observed by the DAMA/LIBRA experiment for the 2$-$6~keV events (where their signal has the highest significance) is 0.0112$\pm$0.0012~dru~\cite{bernabei13}. If we inject \gevcc{10} WIMPs with a 2.96$\times$10$^{-4}$~pb cross section (see in Fig.~\ref{signal_dama} (a)), our detector can observe the same modulation signal as DAMA/LIBRA's.
Based on the Bayesian likelihood fit discussed above, we extract the signal rate by performing 2,000 simulated experiments to obtain a mean of the modulation amplitude as well as 68\%~(1$\sigma$) and 95\%~(2$\sigma$) coverage from 100~kg$\cdot$year to 1050~kg$\cdot$year. Figure~\ref{significance} shows the expected significance for the observation of an annual modulation signal as a function of data size. 
In case of three-year data exposure, corresponding to 600~kg$\cdot$year, the expected significance is 7.2$\pm$1.0~$\sigma$, where the uncertainty means 68\% coverage.
Since DAMA/LIBRA reported a signal significance of 9.3~$\sigma$ with a 1.04~ton$\cdot$years exposure~\cite{bernabei13}, our simulated experiments indicate a similar level of significance as one can see in Fig.~\ref{significance}. 

We repeated the analysis using simulated experiments that assume a 2~dru background level and obtain a significance of 5.1$\pm$0.9~$\sigma$ for the same exposure. In both background scenarios, the median expected significance levels with a three-year data sample are greater than 5$\sigma$, which would be enough to confirm the DAMA/LIBRA observation.  If the KIMS-NaI experiment sees a similar annual modulation signal, we will also identify whether or not the annual modulation events originated from WIMP-induced nuclear recoil signals by taking advantage of relatively good pulse shape discrimination between nuclear recoil and electron recoil events of our NaI(Tl) crystals discussed in Ref.~\cite{Lee:2015iaa}.

\section{Conclusion}
We developed and tested six R\&D stage NaI(Tl) crystals as a part of a program to develop ultra low-background NaI(Tl) crystals for WIMP searches. 
We have reduced the \pbtwoten and \kforty backgrounds in different crystals and achieved a 2~dru background level for recoil energies around 6~keV.
Additional powder purification procedures and the implementation of an LSV system should further reduce the background level to about 1~dru.
We have investigated the expected sensitivity of the KIMS-NaI experiment with a total mass of 200~kg and a three-year period of stable operation
with the assumptions of a 1~dru background rate and a 2~keV energy threshold.
The comparison with the DAMA/LIBRA modulation signal gives confidence that the KIMS-NaI experiment
can test the parameter space claimed by DAMA/LIBRA using the same target material and can achieve similar sensitivity to the reported modulation signatures.
This experiment can unambiguously confirm or refute the DAMA/LIBRA observation.

\section{Acknowledgments}
This research was funded by the Institute for Basic Science (Korea) under project code
IBS-R016-D1 and was supported by the Basic Science Research Program through the National
Research Foundation of Korea funded by the Ministry of Education (NRF-2011-35B-C00007).

\bibliographystyle{plain}

\end{document}